\documentclass[aps,prb,twocolumn,showpacs,superscriptaddress]{revtex4}
\usepackage{epsfig}
\usepackage{graphicx}

\begin{document}

\title{\bf Resonant pairing isotope effect in polaronic systems}

\date{\today}
\author{Julius Ranninger}
\affiliation{Centre de Recherches sur les Tr\`es Basses
Temp\'eratures, Laboratoire Associ\'e \`a l'Universit\'e Joseph
Fourier, Centre National de la Recherche Scientifique, BP 166,
38042, Grenoble C\'edex 9, France}
\author{Alfonso Romano}
\affiliation{Dipartimento di Fisica "E.R. Caianiello",
Universit\`a di Salerno, I-84081 Baronissi (Salerno), Italy -
Unit\`a I.N.F.M. di Salerno}

\begin{abstract}
The intermediate coupling regime in polaronic systems, situated
between the adiabatic and the anti-adiabatic limit, is
characterized by resonant pairing between quasi-free electrons
which is induced by an exchange interaction with localized
bipolarons. The onset of this resonant pairing takes place below a
characteristic temperature $T^*$ and is manifest in the opening of
a pseudogap in the density of states of the electrons. The
variation of $T^*$ is examined here as a function of ($i$) the
typical frequency $\omega_0$ of the local lattice modes, which
determines the binding energy of the bipolarons, and ($ii$) the
total concentration of charge carriers $n_{tot} = n_F + n_B$,
where $n_F$ and $n_B$ are the densities of free electrons and
bipolarons, respectively. The variation of either of these
quantities induces similar changes of the value of $n_B$ with
respect to that of $n_F$, in this way leading to a shift of $T^*$.
For finite, but small values of $n_B$ ($\leq 0.1$ per site), we
find negative and practically doping independent values of the
corresponding isotope coefficient $\alpha^*$. Upon decreasing
$n_{tot}$  such that $n_B$ becomes exponentially small, we find a
rapid change in sign of $\alpha^*$. This is related to the fact
that the system approaches a state which is more BCS-like, where
electron pairing occurs via virtual excitations into bipolaronic
states and where $T^*$ coincides with the onset of
superconductivity.

\end{abstract}

\maketitle

\section{INTRODUCTION}
The experimental verification of an isotope effect in the
classical low temperature superconductors has been an unequivocal
proof for phonon mediated electron pairing in form of Cooper
pairs. Such a clearcut proof is so far absent in the high
temperature superconducting cuprates (HTSC). Moreover, their
proximity in the underdoped regime to an antiferromagnetic
insulating state has largely contributed to conclude that in those
materials pairing is associated with strong correlations.
Nevertheless, the question concerning the origin of the pairing is
far from being settled and lattice driven pairing  ought not to be
ruled out at this stage of our understanding. Actually, there is a
certain amount of experimental evidence that strong
electron-lattice coupling plays some role in stabilizing the
superconducting phase in the cuprates. What is clear is that the
pairing is definitely not of the form of a phonon mediated BCS
one, corresponding to a weak coupling adiabatic regime.

Early on, anomalous mid-infrared optical absorption was found in
virtually everyone of the superconducting cuprates. In lanthanum
based compounds the superconducting critical temperature $T_c$ was
found to scale with the oscillator strength of this
absorption\cite{Orenstein-87,Etemad-88} and later on was shown to
be due to polaronic charge carriers\cite{Eklund-93}. More refined
measurements followed. From neutron scattering studies it became
clear that the manifestation of the superconducting state in the
cuprates could be intimately linked to  strong electron-lattice
coupling. A detailed analysis of the phonon density of states
showed that the high-frequency modes are significantly
renormalized in the superconducting materials  as compared to
their insulating parent compounds\cite{McQueeney-01}. A further
manifestation of strong electron-lattice coupling comes from the
observation of a kink in the electron quasi-particle dispersion in
the 50-80 mev energy region seen in angle resolved photo emission
spectroscopy\cite{Lanzara-01}. Finally, inelastic neutron
scattering experiments pointed to an anomalous behavior  in the
dispersion of the in-plane longitudinal optical phonons with wave
vectors $[0,0.25,0]$ in the YBCO superconductors. This corresponds
to bond stretching  vibrations being associated with dynamical
charge fluctuations on the Cu ions driven by the displacement of
the neighboring ligand environment of the O
atoms\cite{Pintschovius-02}.

Let us suppose, as a working hypothesis, that superconductivity in
the cuprates is indeed controlled by strong electron-lattice
coupling. If we want to test this assumption by examining the
isotope effect in those materials, the right quantity to look at
is  not the transition temperature $T_c$, but the onset
temperature of the electron pairing, $T^*$. $T^*$ shows up in a
qualitative change of the photoemission spectrum such that the
electronic density of states exhibits a charge pseudogap as the
temperature is decreased below $T^*$, eventually merging into a
true superconducting gap below $T_c$. Pair formation in BCS
superconductors coincides with the onset of a global
phase-coherent superfluid state and hence the isotope effect can
be evaluated on the basis of the shift in $T_c$. This does not
apply to the HTSC, where it is a pair resonance state which sets
in below $T^*$, implying pairing on a finite length and time
scale. Only when this length and time scale gets longer and longer
upon decreasing the temperature, a global phase-coherent state can
be established, which is controlled by the center of mass motion
of the Cooper pairs rather than by their breaking up into
individual electron pairs, as in the case of BCS superconductors.
Considering that electron pairing is of resonance type rather than
of a true bound electron pair nature, the different experimental
setups devised to capture such a feature must rely on a time scale
short enough to see this pairing as static. Thus, NMR or NQR
cannot detect it since the relevant time scale has to be well
below $10^{-8}$ sec. On the contrary neutron spectroscopy,
studying the relaxation rate of the crystal field excitations, and
X-ray absorption near edge spectroscopy (XANES) are in the right
time scale regime of $[10^{-13},10^{-15}]$. And in fact it is
those measurements on
La$_{2-x}$Sr$_x$CuO$_4$\cite{Lanzara-99,Furrer-04},
HoBa$_2$Cu$_4$O$_8$\cite{Temprano-00} and
La$_{1.81}$Ho$_{0.04}$Sr$_{0.15}$CuO$_4$\cite{Temprano-02} which
initially demonstrated this resonant pairing isotope effect.

We shall in this paper explore the resonant pairing isotope effect
on the basis of a phenomenological model, believed to capture the
intermediate coupling regime in polaronic systems, situated
between the standard weak coupling adiabatic Born-Oppenheimer
regime (applicable to BCS superconductors) and the anti-adiabatic
regime, where the electrons pair up into bipolarons,
expected to be localized at low temperatures.  We have previously
introduced such a model to study the connection between local
dynamical lattice deformations, measurable by EXAFS pair
distribution functions\cite{Ranninger-02}, and the incoherent
background expected in photoemission spectra, as interpreted in
terms of phonon shake-off processes\cite{Ranninger-98}. On the
basis of this model we shall show that the isotope effect of $T^*$
can be traced back to the pairing energy of the bipolarons which
is a linear function of  the characteristic frequency of the local
lattice modes.

In section II we briefly sketch this model and present a scheme to
determine the doping and frequency dependence of $T^*$. Section
III is devoted to a discussion of the results on the isotope
coefficient and an attempt is made to relate its behavior to the
isotope shift of the bipolaron binding energy and to the
renormalization of the exchange coupling between the localized
bipolaron and itinerant electrons. In the Conclusions, section IV,
we make a comparison of these results with what is known presently
from the the experiments and suggest new ways of looking at this
problem.

\section{The model and resonating pairing temperature}

Exact diagonalization studies\cite{deMello-97} in the crossover
regime between the adiabatic weak coupling limit and the
anti-adiabatic limit have led us to the conjecture that in this
regime we are facing strong fluctuations between tightly bound
pairs and uncorrelated pairs of free electrons, a scenario which
can be phrased into a boson-fermion model. In order to incorporate
into this model the information concerning the origin of the
tightly bound electron pairs, we assume explicitly that they are
of bipolaronic nature. The minimum model which can describe such a
situation is then given by the Hamiltonian
\begin{eqnarray}
& & \!\!\!\!\!\!\!\! H = (D-\mu)\sum_{i,\sigma}n_{i\sigma}
-t\sum_{\langle i\neq j\rangle,\sigma}c^+_{i\sigma}c_{j\sigma}
\qquad \nonumber \\
&&+  (\Delta_B-2\mu) \sum_i \left( \rho_i^z + \frac{1}{2} \right)
+v\sum_i [\rho^+_ic_{i\downarrow}c_{i\uparrow}
  +\rho_i^- c^+_{i\uparrow}c^+_{i\downarrow}] \nonumber \\
&&-  \hbar \omega_0 \alpha \sum_i \left( \rho_i^z + \frac{1}{2}
\right) (a_i+a_i^{+}) +\hbar \omega_0 \sum_i \left(a^{+}_i a_i
+\frac{1}{2}\right). \nonumber \\
\end{eqnarray}
Here $\rho_i^{\pm}$ denote the creation and annihilation operators
for the electron pairs which, due to their interaction with the
local lattice deformations
$X_i=(a_i+a_i^+)/\sqrt{2M\omega_0/\hbar}$, end up in self-trapped
bipolarons $\rho_i^{\pm}\exp\left[\pm \alpha(a_i - a_i^+)\right]$
localized on some effective sites $i$. Such entities are treated
as hard-core bosons with spin-$\frac{1}{2}$ commutation relations,
$\left[ \rho_i^+,\rho_i^-\right]_-=2 \rho_i^z$ and $\left[
\rho_i^+,\rho_j^- \right]_+=\delta_{ij}$. $a_i^{(+)}$ denote
annihilation (creation) operators of the excitations of local
lattice displacements, $M$ is some atomic mass characterizing the
effective sites, and $\omega_0$ is the frequency of the dynamical
local lattice deformations. $c_i^{(+)}$ are the annihilation
(creation) operators for the itinerant electrons with spin
$\sigma$, $n_{i\sigma}=c^+_{i\sigma}c_{i\sigma}$ being the related
number operator. The bare nearest-neighbor hopping integral for
such electrons is given by $t$, corresponding to a bandwidth
$2D=2zt$ where $z$ denotes the lattice coordination number. The
other parameters of the model are the bare bosonic energy level
$\Delta_B$, the coupling $\alpha$ of the local electron pairs to
the surrounding lattice deformations and the bare exchange
coupling $v$ between the bosons and the pairs of itinerant
electrons. The chemical potential $\mu$, being common to electrons
and bosons, guarantees the overall charge conservation. This model
has been studied extensively in the limit of zero coupling
($\alpha=0$) to the lattice, particularly in connection with the
pseudogap effect in the HTSC induced by the resonant pairing of
the electrons. The basic idea behind it is that the closeness of a
weakly bound two-electron state to the energy level of two
itinerant electrons induces resonant pairing with substantial
lifetime in the electronic subsystem. This is an effect which is
analogous to the atom pairing induced by Feshbach resonance in
trapped ultracold gases, studied in connection with their
condensed states\cite{Timmermans-99}.
\begin{figure}
\centerline{\epsfxsize=8cm \epsfbox{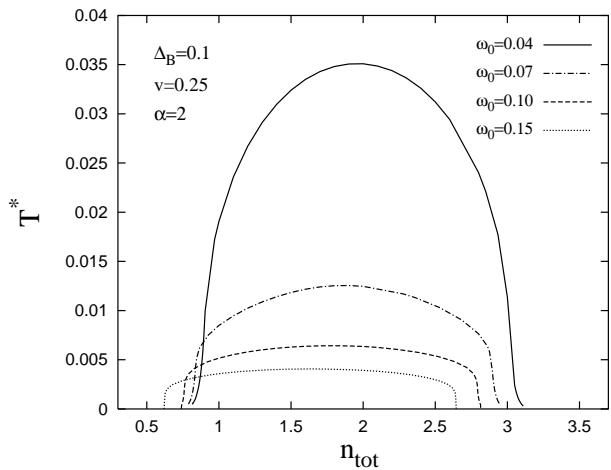}} \caption{$T^*$
as a function of $n_{tot}$ for a variety of different local phonon
frequencies $\omega_0$} \label{T*}
\end{figure}
The onset temperature $T^*$ for electron pairing is determined by
the strong drop-off with decreasing temperature of the on-site
correlation function $\langle \rho_i^+ c_{i\downarrow}
c_{i\uparrow} \rangle$ when passing through
$T=T^*$\cite{Cuoco-03}. The change in this correlation function is
independent on any onset of long-range phase coherence and is
described by mere amplitude fluctuations. Using a variational wave
function of the form
\begin{equation}
\prod_i\left[u(i) + v(i) \rho^+_i \right] |0)\sum_k \left[ u_k
+v_k c^+_{k\uparrow} c^+_{-k\downarrow} \right]|0\rangle \, ,
\end{equation}
we see that the exchange coupling term of the Hamiltonian (1)
becomes
\begin{eqnarray}
v \rho x + vx\sum_i [\rho_i^+ + \rho_i^-] + {v \rho \over 2}\sum_i
[ c_{-k\downarrow}c_{k\uparrow} +
c^+_{k\uparrow}c^+_{-k\downarrow}]
\end{eqnarray}
with
\begin{equation}
x \; = \; {1 \over N} \sum_i \langle c_{i \uparrow}^+ c_{i
\downarrow}^+ \rangle \, , \quad \rho \; = \; {1 \over N} \sum_i
\langle \rho_i^+ + \rho_i^- \rangle
\end{equation}
denoting the amplitudes of the order parameters of the electron
and boson subsystems. Thus, any inter-site phase fluctuations are
explicitly suppressed in such a mean-field approximation. It hence
guarantees that the resulting transition is exclusively due to
amplitude fluctuations and thereby lends itself to describe the
onset of pairing without any simultaneous onset of phase
coherence. That this approach for determining $T^*$ is
qualitatively and, to a large extent, also quantitatively correct,
was checked with a comparative study based on exact
diagonalization procedures\cite{Cuoco-03} and self-consistent
perturbative approaches\cite{Ranninger-03}. The rapid but smooth
drop-off of the local correlation function $\langle \rho_i^+
c_{i\downarrow} c_{i\uparrow} \rangle$ at $T^*$ in those studies
is then apparent as a sharp drop-off to zero of the same function
at a mean-field critical temperature $T^*_{MFA}$. Generally
$T^*_{MFA}$ is found to lie slightly below $T^*$ determined in
more elaborate treatments, but shows the same dependence on the
charge carrier concentration. This justifies the use of an
analogous mean-field type procedure for the generalized
boson-fermion model presented above in Eqs.\,(1-4), in order to
extract a $T^*$ when the coupling of the lattice vibrations to the
charge carriers is turned on.

A detailed account of this mean-field analysis has been given in
ref.\,[11] where we associated our results to a superconducting
phase, assumed to be controlled exclusively by amplitude
fluctuations. The true superconducting phase for this model is
however known to be controlled by phase
fluctuations\cite{Ranninger-03}, while its mean-field phase
describes the pseudogap regime in an approximate form. For the
purpose of the present study we shall adopt such a mean field
analysis for which we shall here merely sketch the procedure.

Since the mean-field decoupling leading to Eq.\,(3) separates the
fermionic part from the bosonic bound electron pairs, we can write
the eigenstates of this Hamiltonian as a direct product of  the
two separate Hilbert spaces associated with fermions and  bosons
in the form
\begin{equation}
|\Psi^{\rm F} \rangle \otimes   \prod_i |\; l \;\}^{B}_i \; ,
\end{equation}
with
\begin{eqnarray}
|\Psi^{\rm F} \rangle &=& \prod_k \left[ u_k + v_k c^+_{k
\uparrow}c^+_{-k \downarrow} \right]| 0 \rangle \; \\
|\; l \;\}^{B}_i &=& \sum_n \left[ u_{l n}(i) +
v_{ln}(i)\,\rho_i^+\right] | 0 )_i|n>_i \; .
\end{eqnarray}
Here $|0\rangle$ and $|0 )$ are the vacuum states for fermions and
bosons, respectively, and $|n\rangle$ is the $n$-th excited
harmonic oscillator state. It should be noted that phonons are
only connected to bosons and thus the contribution (7) is the only
one requiring numerical diagonalization. Denoting the eigenvalues
of the two mean-field states (6) and (7) by
$\tilde\varepsilon_{\bf k}(\rho) = \pm \sqrt{(\varepsilon_{\bf
k}-\mu)^2 + (v \rho)^2/4}$ (which differs from the  bare electron
dispersion $\varepsilon_{\bf k}$ by showing a gap of size $v
\rho$)  and $E_l(x)$, we have the following selfconsistent
equations for the order parameters and the concentration of
electrons and local pairs
\begin{eqnarray}
x &=& -{v \rho \over 4 N} \,\sum_{\bf k} \,{1 \over
\tilde\varepsilon_{\bf k}(\rho)}
\,\tanh{\beta \tilde\varepsilon_{\bf k}(\rho) \over 2} \; ,\\
\rho &=& {1 \over Z}\,
\sum_{ln} \, u_{ln} \, v_{ln} \,
\exp\,\left[-\beta E_l(x)\right] \; ,\\
n_{tot} &=&  \frac{1}{4}\rho^2 + 2 - {1 \over N}\sum_{{\bf k}}
\left( {\varepsilon_{\bf k} \over \tilde\varepsilon_{\bf k}(\rho)}
\, \tanh{\beta \tilde\varepsilon_{\bf k}(\rho) \over 2}\right) \nonumber \\
&+& \frac{1}{Z} \;\sum_{ln} \left[ (u_{ln})^2
- (v_{ln})^2 \right] \exp\,\left[-\beta E_l(x)\right].
\end{eqnarray}
Here $Z=\sum_{l} \exp\,[-\beta E_l(x)]$ denotes  the partition
function corresponding to the bosonic part of the mean-field
Hamiltonian, given by
\begin{eqnarray}
H_B =  (\Delta_B-2\mu) \sum_i \left( \rho_i^z + \frac{1}{2}
\right)
+v x \sum_i [\rho^+_i \; + \rho_i^-] \nonumber \\
-\hbar \omega_0 \alpha \sum_i \left( \rho_i^z + \frac{1}{2}
\right) (a_i+a_i^{+}) + \hbar \omega_0 \sum_i a^{+}_i a_i \; .
\end{eqnarray}
The onset temperature $T^*$ for electron pairing is then
determined by solving the above set of equations in the limit $x
\rightarrow 0$, $\rho \rightarrow 0$.

We assume an electronic band extending from $-D$ to $D$ and choose
a set of parameters $\alpha=2$, $\Delta_B=0.1$, $v=0.25$, with the
phonon frequency $\omega_0$ varying in the range $[0.01,0.1]$,
such as to cover the intermediate polaronic situation (all
energies are in units of the half-bandwidth $D$). This choice,
leading to values of $T^*$ of the order of a few hundred degrees K
($T^* \simeq 10^2 D$), ensures that upon changing the total number
of charge carriers $n_{tot}$, one covers the regime of electron
concentration close to half filling, with the possibility of
having a drastic decrease of $T^*$ with small variations of
$n_{tot}$.

We present in Fig.\,1 the variation of $T^*$ as a function of
$n_{tot}$ for several values of the local phonon frequency
$\omega_0$. As $\omega_0$ is increased, we observe the following
two main effects: ($i$) a shift of the whole curve $T^*(n_{tot})$
to lower values of $n_{tot}$, and ($ii$) an overall diminution of
the value of $T^*$. The first effect is due to a shift $\delta
\Delta_B \simeq \Delta_B - \varepsilon_{BP}$ of the bosonic energy
level associated with the bipolaron binding energy
$\varepsilon_{BP}=\alpha^2 \hbar \omega_0$. The second effect is
due to the decrease of the effective exchange coupling term,
determined by the reduced overlap of the lattice deformations
corresponding to the presence of bipolarons and free electrons,
respectively, on a given site. In the extreme strong coupling
limit $\alpha^2 \hbar \omega_0 \geq D$ this renormalization would
correspond to $v \rightarrow v e^{-\alpha^2}$, but remains of
reasonable size and is frequency dependent  in the intermediate
coupling case. In conclusion, reducing  $n_{tot}$ with doping or
increasing $\omega_0$ upon increasing the isotope mass, leads to
qualitatively similar results in  the shift of $T^*$ as  already
pointed out by some experimental observations\cite{Temprano-00}.

\begin{figure}
\centerline{\epsfxsize=8cm \epsfbox{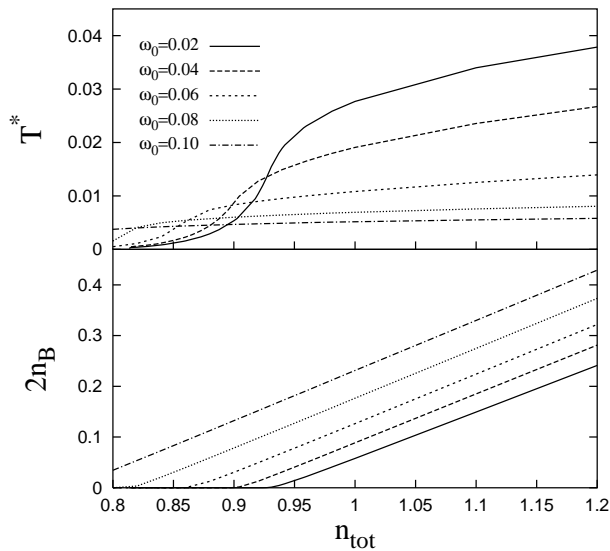}}
\caption{$T^*$ (top panel) and $2n_B$ (bottom pannel) as a
function of $n_{tot}$ for a variety of different local phonon
frequencies $\omega_0$} \label{T*N}
\end{figure}
\begin{figure}
\centerline{\epsfxsize=8cm \epsfbox{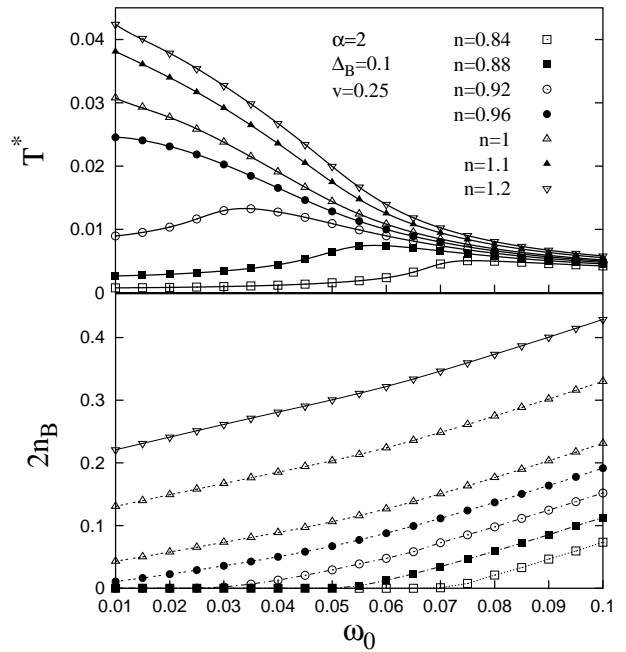}}
\caption{$T^*$ (top panel) and $2n_B$ (bottom panel) as a function
of the local phonon frequency  $\omega_0$ and for a variety of
different $n_{tot}$} \label{T*om}
\end{figure}

\section{the resonating pairing isotope effect}

We shall focus here on a regime of doping ($n_{tot}$) and local
phonon frequency ($\omega_0$) where $T^*$ shows a rapid drop-off
upon decreasing  the total number of charge carriers $n_{tot}$
over a relatively small range of values, i.e.,  $[0.8, 1.2]$. This
choice is made in an attempt to account for the anomalous behavior of
the isotope coefficient $\alpha^*$ observed in the HTSC, which in
the underdoped to optimal doping regime shows unusual and negative
values. In order to show the  evolution of this isotope
effect upon going from the underdoped to the overdoped case (the
latter assumed to be more BCS-like), we examine how $\alpha^*$
changes with the particle density. For doping rates $n_{tot}$ such
that $n_B$ is finite but small we have  resonant pairing of the
electron pairs due to their exchange with localized bipolarons. On
the contrary, if $n_B$ is exponentially small these exchange
processes are purely virtual processes. We then have a situation
where two-particle
pairing can no longer be realized except via a true many-body
effect describing Cooper pairing. The corresponding $T^*$ then
signals simultaneous pairing and onset of superconductivity
together with an isotope exponent which is positive.

The behavior of $T^*$ as a function of $n_{tot}$ and $\omega_0$ is
controlled by two competing effects:

($i$) $T^*$ increases with increasing $n_{tot}$ when $n_B$ varies
between 0 and 0.5.

($ii$) $T^*$, for a fixed $n_B$, decreases with increasing
$\omega_0$ because of the polaron induced reduction of the
exchange coupling $v$. But since an increase of $\omega_0$  not only
reduces the effective exchange coupling  but also leads to an
increased bipolaron level shift which in turn increases $n_B$ for
a fixed $n_{tot}$, $T^*$ is in general a non-monotonic function of
$\omega_0$.

We present in Fig.\,2 the variation of $T^*$ and $n_B$ as a
function of $n_{tot}$  in the above mentioned regime of
parameters. We notice the onset of a rapid rise of $T^*$ with
increasing $n_{tot}$ controlled by an equally rapid rise of $n_B$.
Upon further increasing $n_{tot}$, $T^*$ starts to saturate, an
effect due to the hard-core nature of the localized bipolarons
which becomes important when $n_B$ approaches $0.5$.

The variation of $T^*$ and $n_B$ with $\omega_0$ for a set of
different $n_{tot}$ is illustrated in Fig.\,3. For sufficiently
large values of $n_{tot}$, such that $n_B$ is finite but still
lower than 0.5, we find a monotonically decreasing function with
increasing $\omega_0$, controlled  by the polaron induced
reduction of the effective exchange coupling v. As $n_{tot}$
decreases, leading to a vanishing concentration of bipolarons, the
behavior of $T^*$ changes qualitatively. For low values of
$\omega_0$, the polaron induced reduction of the effective
exchange coupling is negligible and $T^*$ now increases with
increasing $\omega_0$ because it is controlled by the increase of
$n_B$. With increasing $\omega_0$, the effect of the polaron
induced reduction of the exchange coupling becomes competitive
with the increase of $n_B$ such that the initial increase of $T^*$ with
increasing $\omega_0$ changes into a decreasing behavior.

Provided we are in the regime of resonant pairing, with $n_B$
small but finite and $T^*$ monotonically decreasing with
decreasing $n_{tot}$, Fig. 3 tells us the following. $T^*$ is
shifted upwards by the decrease of the phonon frequency associated
with the increase of the isotope mass (as realized, for instance,
replacing $^{16}$O by $^{18}$O), and this effect becomes less and
less pronounced as doping increases ($n_{tot}$ decreases).
Moreover, if for a given doping level the isotope substitution is
made for heavier elements, such as $^{63}$Cu replaced by
$^{65}$Cu, the increase of the corresponding $T^*$ is getting
smaller and smaller, as one can deduce from the behavior of $T^*$
at low $\omega_0$. These features are in qualitative agreement
with experimental findings in LaSrHoCuO$_4$
compounds\cite{Furrer-04,Temprano-02}.

\begin{figure}
\centerline{\epsfxsize=8cm \epsfbox{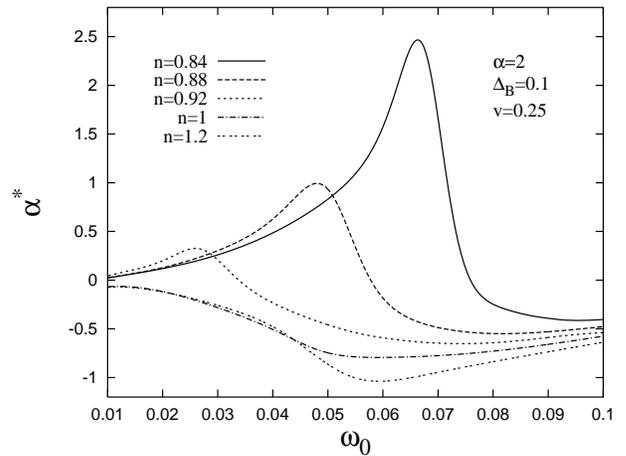}}
\caption{$\alpha^*$ as a function of the local phonon frequency
$\omega_0$ and for a variety of different $n_{tot}$}
\label{alpha*om}
\end{figure}

\begin{figure}
\centerline{\epsfxsize=8cm \epsfbox{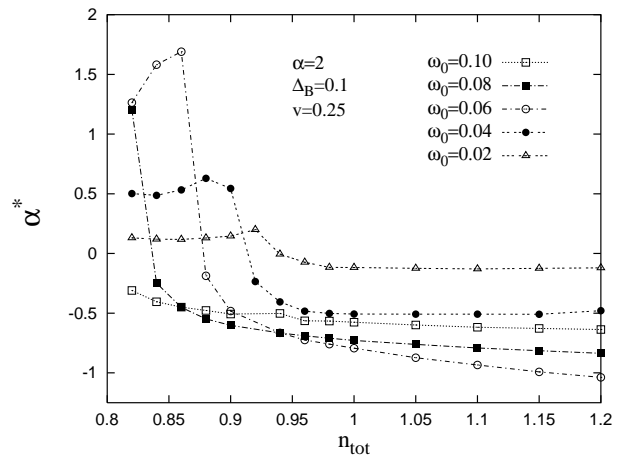}}
\caption{$\alpha^*$ as a function of $n_{tot}$ and for a variety
of different local phonon frequencies  $\omega_0$.}
\label{alpha*n}
\end{figure}

In order to determine the isotope coefficient, which itself will
depend on $\omega_0$ and $n_{tot}$, we have first interpolated the
calculated values of $T^*$  by a ratio of polynomials, and then derived
$\alpha^*$ from the relation
\begin{equation}
\alpha^* = 0.5 \; \ln \left({T^*(i) \over T^*(i+1)}  /
{\omega_0(i) \over \omega_0(i+1)} \right)
\end{equation}
where $\omega_0(i)$ and  $\omega_0(i+1)$ represent two very close
values of the phonon frequency. We present in Fig.\,\ref{alpha*om}
the variation of the isotope exponent as a function of the phonon
frequency $\omega_0$ for a set of different values of $n_{tot}$
and in Fig.\,\ref{alpha*n} the variation of the isotope exponent
as a function of $n_{tot}$ for a selected set of phonon
frequencies $\omega_0$.

We notice two distinct regimes which characterize $\alpha^*$:

($i$) a regime where $\alpha^*$ is negative and depends strongly on
$\omega_0$ but shows a relative independence on the concentration
$n_{tot}$. This happens when the corresponding $n_B$ is
big enough to sustain the mechanism of resonant pairing.

($ii$) a regime of positive values of $\alpha^*$ which occurs
when $n_B$ drops to zero as a consequence of the
decrease of $n_{tot}$ or, alternatively, of $\omega_0$ (see the
bottom panel of Fig.\,2). In this case, which corresponds to the
system being more BCS-like, electron pairing arises from virtual
excitations of the electrons into bipolaronic localized states
above the Fermi level and the onset of pairing at $T^*$ coincides
with the onset of superconductivity.

\begin{figure}
\centerline{\epsfxsize=8cm \epsfbox{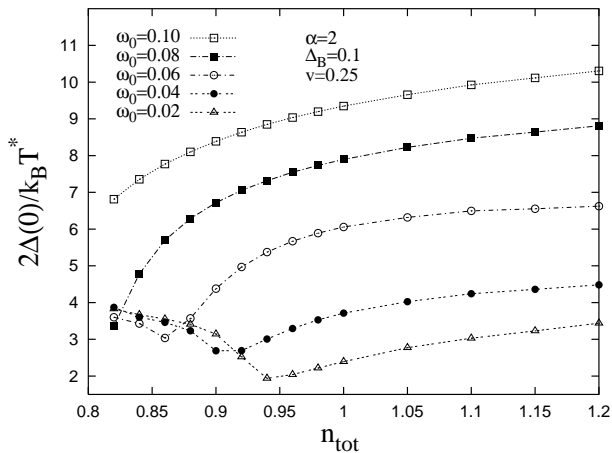}} \caption{$2
\Delta(0) /(k_B T^*)$ as a function of $n_{tot}$  and for a
variety of different local phonon frequencies  $\omega_0$.}
\label{gapratio}
\end{figure}

These two regimes are clearly visible in the variation of the gap
ratio $2 \Delta(0) /(k_B T^*)$, where $2 \Delta(0) = v\rho(0)$ is
the zero temperature gap in the Fermionic excitation spectrum
$\tilde{\varepsilon}_k$. This ratio, which is shown in
Fig.\,\ref{gapratio} as a function of $n_{tot}$ for several
$\omega_0$, is a slowly-varying function of the concentration when
the system is well inside the resonant pairing regime ($1.2 <
n_{tot} < 2.8$ for our choice of $\omega_0$ values) but strongly
depends on the local phonon frequency $\omega_0$. For values of
$n_{tot}$ such that $n_B$ becomes exponentially small (depending
on $\omega_0$ this happens below $n_{tot}=0.9$, see bottom panel
in Fig.\,2), the gap ratio approaches the BCS value 3.52.

Going back to the behavior of the isotope coefficient at low
$n_{tot}$, we stress that even in this BCS-like regime $\alpha^*$
shows a frequency dependence related to the pairing mechanism,
which in the case studied here of strong electron-phonon
interaction is different from the standard weak-coupling BCS one.
In our model, pairing among the electrons is induced via virtual
excitations of electron pairs into a localized bipolaron state and
an effective BCS-like Hamiltonian for this situation can be
derived along the following two steps:

i) the boson-phonon coupling is incorporated into an effective
boson-fermion exchange interaction $\tilde v =
v\exp{(-\alpha^2/2)}$ via a Lang-Firsov approximation;

ii) the boson-fermion coupling term is subsequently eliminated to
linear order via the usual unitary transformation
\begin{eqnarray}
\tilde H &=& e^S H e^{-S}\\
S &=& \sum_{i,k} f(k)[\rho_i^+ c_{-k\downarrow}c_{k\uparrow}
-{\rm H.c.}] \\
f(k) &=& {\tilde v \over 2\varepsilon_k-\Delta_B +
\varepsilon_{BP}} \label{EffCoup}
\end{eqnarray}
which in the end results in an effective BCS-like Hamiltonian of
the form
\begin{eqnarray}
\tilde H &=& \sum_{k,\sigma} (\varepsilon_k - \mu)c^+_{k\sigma}
c_{k\sigma} \nonumber\\
&& +\, \frac{{\tilde v}^2}{2}\,\sum_{k,k'}\,\left[f(k) +
f(k')\right] c^+_{k\uparrow} c^+_{-k\downarrow}
c_{-k'\downarrow}c_{k'\uparrow} \; . \label{EffHam}
\end{eqnarray}

In the standard weak-coupling effective BCS Hamiltonian, the
interaction term is restricted to the summation over $k$-vectors
within a small energy range around the Fermi surface of width
given by the Debye temperature $\Theta_D$. This gives rise to an
effective interaction of the order of $\Theta_D$ from which a
frequency independent isotope shift follows, with an isotope
exponent equal to 0.5.

The effective Hamiltonian for our scenario, eq.\,\ref{EffHam},
shows on the contrary no such a cutoff in energy and extends the
pairing of electrons over all $k$ vectors below the Fermi vector,
albeit with different weight. Moreover, because of the presence of
the bipolaronic energy in the denominator of eq.\,\ref{EffCoup},
the effective interaction is attractive in this regime of low
bipolaron concentration, since $2 \varepsilon_{k_F} < \Delta_B -
\varepsilon_{BP}$, and grows in magnitude as $\omega_0$ increases.
This is the reason why we obtain a frequency dependent isotope
shift even in this BCS-like regime with a $T^*$ which increases as
$\omega_0$ increases.

\section{Conclusions}

Experimentally, there are strong indications that in the HTSC the
isotope coefficient associated with the temperature $T^*$ at which
the pseudogap in the underdoped regime opens up, is negative. Its
precise numerical value depends on the type of  material, the
doping regime, the type of isotope substitution ($^{16}$O
$\leftrightarrow\, ^{18}$O\, ,$\; ^{63}$Cu $\leftrightarrow\,
^{65}$Cu) as well as on the different time scales of the various
experiments. Presently, there exist to our knowledge no systematic
experimental studies which would permit to test a particular
theoretical approach on this issue in detail. The study presented
here was designed to incite experimental work to explore
specifically the doping dependence of the resonant pairing isotope
effect, given the characteristic strong doping dependence of $T^*$
in the HTSC. If we assume that this feature is related to a
two-component scenario with localized charge carriers (such as
bipolarons) and itinerant electrons, with the sharp drop of $T^*$
being essentially governed by a doping dependent change of the
bipolaron thermal population, the isotope effect on $T^*$ should
exhibit a corresponding concentration dependence. In agreement
with experiments, we find indeed  a negative value for the isotope
exponent, as long as pairing is assured by a resonant scattering
between the localized bipolarons and the free electrons. However,
as soon as upon doping we enter a regime where the bipolaron level
moves above the Fermi energy, pairing is only possible via a
collective effect such as Cooper pairing and the isotope  exponent
switches sign and becomes positive. With respect to the HTSC, this
could happen when going from the underdoped into the overdoped
regime. Considering, however, that the lattice mediated coupling
between electrons in the present scenario is different from the
standard phonon-induced Cooper pairing, we obtain a behavior for
the isotope exponent which deviates from that of standard low
temperature BCS superconductors. Although it converges to a value
independent on doping as $n_{tot}$ is reduced such that $n_B$
becomes exponentially small (see the curves at low $\omega_0$ in
Fig.\,5), it nevertheless continues to sensitively depend on the
characteristic local phonon frequency $\omega_0$.

The present mean-field type study is expected to qualitatively
correctly describe the doping dependence of $\alpha^*$ in the
resonant pairing regime where $T^*$ is controlled by amplitude
fluctuations. This mean-field scheme treats the local dynamical
atomic displacements as correlated to the local density
fluctuations between the electrons and the bipolarons, which in
our model, are responsible for the opening of the pseudogap.  It
is this mechanism which is at the origin of the blockage of the
crystal field excitations which below $T^*$ couple to the free
charge carriers, as observed experimentally in neutron
spectroscopic measurements testing the transitions between
different crystal field levels\cite{Temprano-00,Temprano-02}. A
more quantitative analysis than the mean-field procedure presented
here would be required in order to account for the dynamical
nature of the onset of pairing, as seen in experiments with
different time scales\cite{Zhao-02} which lead to different
absolute values of $T^*$ but to qualitatively similar doping
dependence of $T^*$. Such a highly non-trivial undertaking is
however beyond the scope and purpose of the present work.

We want to stress that the scenario described here is based on a
mechanism originating from {\it purely local} dynamical lattice
instabilities for which experimental evidence has been
accumulating over the past few years. Inelastic neutron scattering
measurements have shown strong compositional dependence of certain
optical (half-breathing zone-edge) phonon modes, which were linked
to spatial local charge inhomogeneities\cite{Chung-03} of small
clusters and suggest that the lattice is strongly involved in the
charge dynamics. EXAFS studies\cite{Roehler-04} tracked such local
charge inhomogeneities in form of a significant deviation from a
systematic Pauling-type shift of the planar Cu--O bonds, when the
bonding mechanism changes from ionic to more covalent nature as
doping is increased. Similarly, very recent tunneling
spectroscopic studies\cite{Davis-04} indicate the existence of
local charge density modulations involving local spatial
correlations of four CuO$_2$ unit cells, referred to as "squared
checkerboard" structures. All these findings go in the direction
of {\it local} charge inhomogeneities as well as {\it local}
dynamical lattice deformations, against earlier propositions of
long-range stripe order, on the basis of which the pseudogap
isotope effect was theoretically investigated
previously\cite{Andergassen-01}.


\begin{thebibliography}{99}
\bibitem{Orenstein-87} J.~Orenstein, G.~A.~Thomas, D.~H.~Rapkine, C.~G.~Bethea,
B.~F.~Levine, B.~Batlogg, R.~J.~Cava, D.~W.~Johnson, Jr., and
E.~A.~Rietman, Phys. Rev. B {\bf 36}, 8892 (1987).
\bibitem{Etemad-88} S.~Etemad, D.~E.~Aspnes, M.~K.~Kelly, R.~Thompson,
J.~M.~Tarascon, and G.~W.~Hull, Phys. Rev. B {\bf 37}, 3396
(1988).
\bibitem{Eklund-93} X.~X.~Bi and P.~C.~Eklund, Phys. Rev. Lett. {\bf 70}, 2625
(1993).
\bibitem{McQueeney-01} R.~J.~McQueeney, J.~L.~Sarrao, P.~G.~Pagliuso,
P.~W.~Stephens, and R.~Osborn, Phys. Rev. Lett. {\bf 87}, 077001
(2001).
\bibitem{Lanzara-01} A.~Lanzara, P.~V.~Bogdanov, X.~J.~Zhou, S.~A.~Kellar,
D.~L.~Feng,
E.~D.~Lu, T.~Yoshida, H.~Eisaki, A.~Fujimori, K.~Kishio, J.-I.
Shimoyama, T.~Noda, S.~Uchida, Z.~Hussain, and Z.-X.~Shen, Nature
(London) {\bf 412}, 510 (2001).
\bibitem{Pintschovius-02} L.~Pintschovius, W.~Reichardt, M.~Kl\" aser,
T.~Wolf, and H.~v.~L\" ohneysen, Phys. Rev. Lett. {\bf 89}, 037001
(2002).
\bibitem{Lanzara-99} A.~Lanzara,  Guo-meng Zhao, N.~L.~Saini, A.~Bianconi,
K.~Conder, H.~Keller, and K.~A.~M\"uller, J. Phys.:  Cond. Matt.
{\bf 11}, L541 (1999).
\bibitem{Temprano-00} D.~Rubio Temprano, J.~Mesot, S.~Janssen,
A.~Furrer, K.~Conder, and H.~Mutka, Phys. Rev. Lett. {\bf 84},
1990 (2000).
\bibitem{Temprano-02} D.~Rubio Temprano, K.~Conder, A.~Furrer, H.~Mutka,
V.~Trounov, and K.~A.~M\" uller, Phys. Rev. B {\bf 66}, 184506
(2002).
\bibitem{Furrer-04} A.~Furrer, K.~Conder, P.~H\"afliger, and A.~Podlesnyak,
Physica C {\bf 408}, 773 (2004).
\bibitem{Ranninger-02} J.~Ranninger and A.~Romano, Phys. Rev. B {\bf 66},
094508 (2002).
\bibitem{Ranninger-98} J.~Ranninger and A.~Romano, Phys. Rev. Lett. {\bf 80},
5643 (1998).
\bibitem{deMello-97} E.~V.~L.~de Mello and J.~Ranninger, Phys. Rev. B {\bf 55},
14872 (1997).
\bibitem{Timmermans-99} E.~Timmermans, P.~Tommasini, M.~Hussein, and A.~Kerman,
Phys. Rep. {\bf 315}, 199 (1999).
\bibitem{Cuoco-03} M.~Cuoco, C.~Noce, J.~Ranninger, and A.~Romano, Phys. Rev. B
{\bf 67}, 224504 (2003).
\bibitem{Ranninger-03} J.~Ranninger and L.~Tripodi, Phys. Rev. B {\bf 67},
174521 (2003).
\bibitem{Zhao-02} Guo-meng Zhao, H.~Keller, and K.~Conder, J. Phys.: Cond.
Matt. {\bf 13}, R569 (2001).
\bibitem{Chung-03} J.-H.~Chung,  T.~Egami, R.J.~McQueeney,
M.~Yethiraj, M.~Arai, T.~Yokoo, Y.~Petrov, H.A.~Mook, Y.~Endoh,
S.~Tajima, C.~Frost, and F.~Dogan, Phys. Rev. B {\bf 67}, 014517
(2003)
\bibitem{Roehler-04} J.~Roehler, cond-mat/0407654 (to be published in Int.
J. Mod. Phys. B).
\bibitem{Davis-04} T.~Hanaguri, C.~Lupien, Y.~Kohsaka, D.H.~Lee, M.~Azuma,
M.~Takano, H.~Takagi, and J.C.~Davis, Nature {\bf 430}, 1001
(2004).
\bibitem{Andergassen-01} S.~Andergassen, S.~Caprara, C.~Di Castro, and
M.~Grilli, Phys. Rev. Lett. {\bf 87}, 56401 (2001).
\end{thebibliography}
\end{document}